\documentclass{phbauth}        
\usepackage{graphicx}

\begin{document}

\begin{frontmatter}

\title{Monolayers of $^3$He on the surface of bulk superfluid $^4$He}

\author[address1]{Jonathan\,P.\,Warren} and
\author[address1]{Charles\,D.H.\,Williams\thanksref{thank1}},

\address[address1]{School of Physics, University of Exeter, Stocker Road,
Exeter EX4 4QL, United Kingdom}

\thanks[thank1]{Corresponding author. E-mail: c.d.h.williams@ex.ac.uk} 

\begin{abstract}
We have used quantum evaporation to investigate the two-dimensional fermion system that forms at the free surface of (initially isotopically pure) $^4$He when small quantities of $^3$He are added to it. By measuring the first-arrival times of the evaporated atoms, we have determined that the $^3$He--$^3$He potential in this system is $V_\mathrm{3S}/k_\mathrm{B}=0.23\pm0.02\,\mathrm{ K\,nm^2}$ (repulsive) and estimated a value of $m_\mathrm{3S}=(1.53\pm0.02)m_\mathrm{3}$ for the zero-coverage effective mass. We have also observed the predicted second layer-state which becomes occupied once the first layer-state density exceeds about 0.6 monolayers.
\end{abstract}


\end{frontmatter}



When small quantities of $^3$He are added to bulk superfluid $^4$He below $T{\sim}100{\rm\, mK}$ the atoms occupy so-called Andreev states \cite{ANDREEV} and form a degenerate two-dimensional fermion system. At low temperatures, and when the surface density of $^3$He $n_\mathrm{3S}$ is less than about half a monolayer, the $^3$He chemical potential can be described \cite{GUO} by
\begin{equation} 
\mu_3=E^{(0)}_\mathrm{3S}+\left(\frac{\pi\hbar^2}{m_\mathrm{3S}}+\frac{ V_\mathrm{3S}}{2}\right)n_\mathrm{3S}
\label{mu3}
\end{equation}
where $E^{(0)}_\mathrm{3S}$ is the binding energy of a single $^3$He atom (effective mass $m_\mathrm{3S}$) to the $^4$He surface. $V_\mathrm{3S}$ parameterises the $^3$He--$^3$He interaction.
\par
Values of $m_\mathrm{3S}$ and $V_\mathrm{3S}$ inferred from measurements of thermodynamic properties of the surface, such as surface tension, are strongly covariant (see \cite{EDWARDS} for example). However, we have been able to obtain independent values for these quantities using a new method based on quantum evaporation \cite{BAIRD,CDHW}, as follows:
\begin{figure}[b]
\begin{center}\leavevmode
\includegraphics[width=1.0\linewidth]{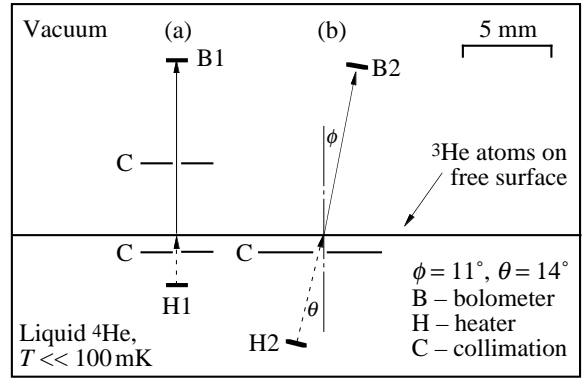}
\caption{Schematic diagram of the quantum evaporation experiment} \label{EXPT}\end{center}\end{figure}
\par
A tightly collimated beam of high-energy phonons \cite{PHONONS} was directed at normal incidence to the free liquid surface (Fig.\ \ref{EXPT}a). The evaporated $^3$He atom beam was also collimated so that only atoms with a nearly zero component of momentum parallel to the surface were detected. An incident phonon of energy $E_\mathrm{p}$ imparts kinetic energy to the ejected atom, which has bare mass $m_3$, such that
\begin{equation} 
{\frac{1}{2}} m_3v^2-E_\mathrm{p}=E^{(0)}_\mathrm{3S}+{\frac{1}{2}}V_\mathrm{3S} n_\mathrm{3S}.
\label{KE}
\end{equation}

\begin{figure}[b]
\begin{center}\leavevmode
\includegraphics[width=1.0\linewidth]{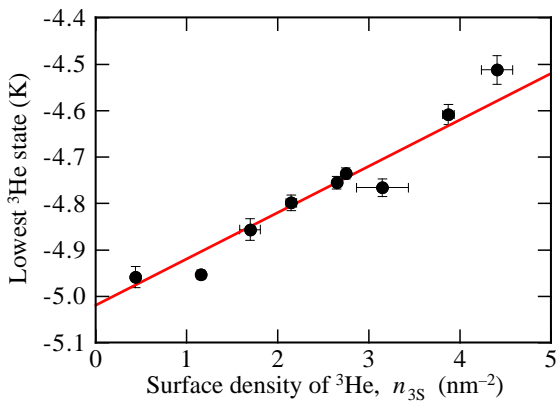}
\caption{Values of $E^{(0)}_\mathrm{3S}+\frac{1}{2}V_\mathrm{3S} n_\mathrm{3S}$ deduced from measured first-arrival times for $^3$He atoms evaporated by high-energy phonons as a function of surface density $n_\mathrm{3S}$.} \label{ENERGIES}\end{center}\end{figure}

The distribution of phonon energies arriving at the surface is unaffected by $n_\mathrm{3S}$ at the coverages used, and the value of $E^{(0)}_\mathrm{3S}/k_{\mathrm B}=-5.02\pm0.03\,\mathrm{K}$ \cite{EDWARDS} is constant. Therefore, the variation with $n_\mathrm{3S}$ of arrival times of $^3$He atoms at the bolometer is due simply to the term $\frac{1}{2}V_\mathrm{3S} n_\mathrm{3S}$ in Eqn\ \ref{KE}. Our measured first-arrival times were consistent with Eqn\ \ref{KE} for coverages up to  $n_\mathrm{3S}=4\,\mathrm{nm}^{-2}$ (Fig.\ \ref{ENERGIES}) and we conclude that
$V_\mathrm{3S}/k_{\mathrm B}=(0.23\pm0.02)\,\mathrm{K\,nm^2}$. Knowledge of this value eliminates the uncertainty due to covariance (see above) in the value of $m_\mathrm{3S}$ inferred from measurements \cite{EDWARDS,EDWARDS2} of surface-sound velocity and surface tension. Hence, the best estimate of $m_\mathrm{3S}$ can refined from $(1.45\pm0.10)m_\mathrm{3}$ to $m_\mathrm{3S}=(1.53\pm0.02)m_\mathrm{3}$.

\par
\begin{figure}[btp]
\begin{center}\leavevmode
\includegraphics[width=1.0\linewidth]{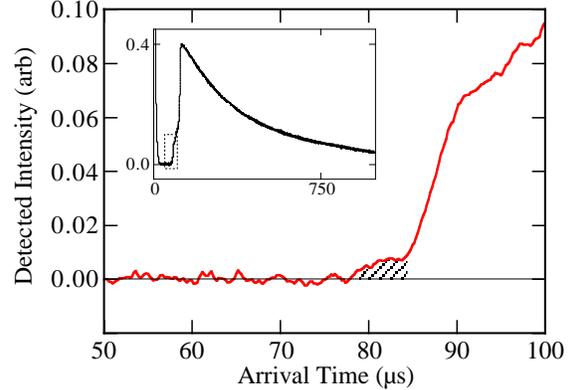}
\caption{$^3$He atoms from the lowest surface-state start arriving at $85\,\mathrm{\mu s}$, the earlier component (shaded) is due to atoms from the excited surface state. The largest signal component (inset) is due to evaporated $^4$He atoms.} \label{SECONDSTATE}\end{center}\end{figure}

We have also used a slight variant of the experiment (Fig.\ \ref{EXPT}b) to search for evidence of a second `excited' state predicted by Pavloff and Treiner \cite{PAVLOFF}. Atoms in this state have a smaller binding energy $E^{(1)}_\mathrm{3S}$ to the surface than those in the first surface state, and the signature of its occupation is therefore an additional faster component in the detected signal. This component appeared, albeit at a level not greatly above the detector noise, on all signals taken below $T=60\,\mathrm{mK}$ with surface coverages of, and above, about 0.6 monolayers, {\it i.e.} $n_\mathrm{3S}=4\,\mathrm{nm}^{-2}$ (Fig.\ \ref{SECONDSTATE}). From our preliminary measurements of first-arrival times, we find that $E^{(1)}_\mathrm{3S}=-3.4\pm0.4\,\mathrm{K}$, in agreement with the predictions \cite{PAVLOFF}.
\par
Although this paper has discussed exclusively the states of $^3$He above bulk $^4$He, we note that thin film and layered systems have some comparable properties and have been investigated by other groups using NMR, third-sound and heat capacity measurements \cite{ANDERSON,DANN,MORISHITA}.

\end{document}